\documentstyle[aps,prl,floats,amssymb]{revtex}
\def\IC{\bf C}
\def\IZ{\bf Z}

\def\z2z2{$\IC^3/(\IZ_2\times\IZ_2)$}

\def\id{{\bf 1}}

\def\th{\theta}

\def\beq{\begin{equation}}\def\eeq{\end{equation}}
\def\beqa{\begin{eqnarray}}\def\eeqa{\end{eqnarray}}
\def\barr{\begin{array}}\def\earr{\end{array}}

 \let\br=\bigr

\def\bd{\begin{document}}
\def\ed{\end{document}}
\def\ba{\begin{array}}
\def\ea{\end{array}}
\def\bea{\begin{eqnarray}}
\def\eea{\end{eqnarray}}
\def\ft#1#2{{\textstyle{{\scriptstyle #1}\over {\scriptstyle #2}}}}
\def\fft#1#2{{#1 \over #2}}
\newcommand{\be}{\begin{equation}}
\newcommand{\ee}{\end{equation}}
\newcommand{\eq}[1]{(\ref{#1})}
\def\eqs#1#2{(\ref{#1}-\ref{#2})}
\def\det{{\rm det\,}}
\def\tr{{\rm tr}}
\newcommand{\ho}[1]{$\, ^{#1}$}
\newcommand{\hoch}[1]{$\, ^{#1}$}
\def\ra{\rightarrow}

\def\Xh{\hat{X}}
\def\ah{\hat{a}}
\def\xh{\hat{x}}
\def\yh{\hat{y}}
\def\ph{\hat{p}}
\def\G{{\cal G}}
\def\Dth{{\Delta_\th}}

\def\bk{{\bf k}}
\def\bx{{\bf x}}
\def\br{{\bf r}}
\def\tr{{\rm tr \,}}
\def\Tr{{\rm Tr \,}}
\def\diag{{\rm diag \,}}
\def\tg{{\rm tg \,}}
\def\NPB#1#2#3{Nucl. Phys. B {\bf #1} (19#2) #3}
\def\PLB#1#2#3{Phys. Lett. B {\bf #1} (19#2) #3}
\def\PLBold#1#2#3{Phys. Lett. {#1B} (19#2) #3}
\def\PRD#1#2#3{Phys. Rev. D {\bf #1} (19#2) #3}
\def\PRL#1#2#3{Phys. Rev. Lett. {\bf #1} (19#2) #3}
\def\PRT#1#2#3{Phys. Rep. {\bf #1} C (19#2) #3}
\def\MODA#1#2#3{Mod. Phys. Lett.  {\bf #1} (19#2) #3}
\def\ov{\overline}

\begin{document}

\preprint{UPR-948-T, CERN-TH/2001-190}
\twocolumn[\hsize\textwidth\columnwidth\hsize\csname@twocolumnfalse\endcsname

\title{Three-Family Supersymmetric Standard-like Models  \\
from Intersecting Brane Worlds}

\author{Mirjam Cveti\v c$^{1,2}$,  Gary Shiu$^{1,2}$ and Angel M. Uranga$^2$} 
\address{$^1$Department of Physics and Astronomy, 
University of Pennsylvania, Philadelphia PA 19104-6396, USA\\
$^2$  Theory Division, CERN, CH-1211 Geneva 23, Switzerland}

\maketitle

{\tighten
\begin{abstract}
We construct the first three family $N=1$ supersymmetric string model with 
Standard Model gauge group $SU(3)_C\times SU(2)_L\times U(1)_Y$ from an
orientifold of type IIA theory on ${\bf T}^6/({\bf Z}_2\times {\bf Z}_2)$ 
and D6-branes intersecting at angles. In addition to the  minimal
supersymmetric Standard Model particles, the model contains right-handed
neutrinos, a chiral (but anomaly-free) set of exotic multiplets, and
extra vector-like multiplets. We discuss some phenomenological features
of this model. 
\end{abstract}
}
]\narrowtext 

The space of classical string vacua is highly degenerate, and at present 
we  are unable  to make definitive statements about how the string vacuum
describing our universe is selected. Nonetheless, one can use
phenomenological constraints as guidelines to construct semi-realistic
string models and explore, with judicious assumptions, the resulting
phenomenology. The purpose of such explorations is of course not to find
{\it the} model which would  fully describe our world, but to examine 
the generic features of these string derived solutions.
 
Until a few years ago, such explorations were carried out mainly in the
framework of weakly coupled heterotic string theory. Indeed, a number of 
semi-realistic string models have been constructed and analyzed \cite{review}.
However, an important lesson from string duality is that these models
represent only a corner of  M-theory -- the string vacuum describing
our world may well
be in a completely different regime in which the perturbative description
of heterotic string theory breaks down \cite{burt}.
Fortunately, the advent of D-branes allows for the 
construction of semi-realistic
string models in another calculable regime, as illustrated by the various
four-dimensional $N=1$ supersymmetric Type II orientifolds 
($\!$\cite{ABPSS,berkooz,N1orientifolds,zwart,ShiuTye,afiv,wlm,CPW,kr,CUW} 
and references therein) constructed using conformal field theory
techniques.
However, the constraints on supersymmetric four-dimensional models are
rather restrictive, leading to not fully realistic gauge sectors and
matter contents. Motivated by the search for Standard Model-like
solutions, several discrete or continuous deformations of this class of
models have been explored.
They include:
(i) blowing-up of orientifold singularities \cite{clew,pru}, which
cannot be  described by a free world-sheet conformal field theory,
hence the spacetime spectrum can only be computed using field theory
techniques, (ii) locating the branes at different points in the internal
space (see e.g. \cite{wlm,CUW,cl}) which in a T-dual picture corresponds
to turning on continuous or discrete Wilson lines, (iii) introduction of
discrete values for the NS-NS $B$ field \cite{ShiuTye,sagnottietal} which
in the T-dual picture corresponds to tilting the compactification tori,  
(iv) introduction of gauge fluxes in the D-brane world-volumes (see
\cite{bachas} for an earlier discussion, and 
\cite{magnetised,bkl} for supersymmetric $D=6$ models), which in the  
T-dual version  corresponds to  D-branes intersecting at angles (hence
closely related to models in \cite{blumen,bonn}). 
 
An appealing feature of (iv) is that generically, there exists chiral
fermions where D-branes intersect \cite{bdl}. Their multiplicity is hence
determined by a topological quantity, {\it i.e.} the intersection number
of the branes. Models with D-branes intersecting at arbitrary angles are
non-supersymmetric. For non-supersymmetric models, Ramond-Ramond (RR)
tadpole cancellation conditions are less constraining \cite{ads-au}, a
fact exploited to construct semi-realistic models in \cite{aiq,aiqu}, and
more recently in \cite{bkl,bonn,bgkl,afiru,imr} in the context of
intersecting branes. However, non-supersymmetric models suffer from 
a less understood, complicated dynamics. At the quantum level, flat
directions are lifted, leading to involved stabilization problems. In addition
uncancelled NS-NS tadpoles require redefining the background geometry
\cite{nsns}. These difficulties are the main reasons that we focus on the
construction of string models with $N=1$ supersymmetry.

The purpose of this letter is to present the first example of a
four-dimensional $N=1$ supersymmetric type IIA orientifold with D6-branes
intersecting at angles, leading to Standard Model gauge group (as part of 
the gauge group structure) and three quark-lepton families. Beyond the 
structure of the  minimal supersymmetric Standard Model (MSSM), the model
contains some additional gauge factors, right-handed neutrinos, a chiral
set of fields with exotic standard model gauge quantum numbers, and
diverse vector-like multiplets. Despite its lack of fully realistic
features, it provides the first construction of supersymmetric Standard
Model-like string models in the setup of (non-trivial) intersecting brane
worlds. Interestingly, since only D6-branes and O6-planes are involved,
the M-theory lift of this general class of supersymmetric orientifold 
models corresponds to purely geometrical backgrounds admitting a $G_2$
holonomy metric and leading to chiral four-dimensional fermions.

We shall provide the key features of the construction, the gauge group
structure and the massless spectrum. The details of the construction,
consistency condition,s as well as a broader class of models (including 
examples of grand unified theories) will be presented in a longer
companion paper \cite{CSU}.

The construction of the model is based on an orientifold of type IIA on
${\bf T}^6/({\bf Z}_2\times {\bf Z}_2)$ (related to \cite{berkooz} by
T-duality), with D6-branes not parallel to
the orientifold 6-planes (O6-planes). The generators $\theta$, $\omega$ 
act as $ \theta:   (z_1,z_2,z_3) \to (-z_1,-z_2,z_3)$, and  $\omega:
(z_1,z_2,z_3) \to (z_1,-z_2,-z_3)$ on the complex coordinates $z_i$ of 
${\bf T}^6$, which we moreover choose to be factorizable. The orientifold
action  is $\Omega R$, where $\Omega$ is world-sheet parity, and $R$ acts
by $ R:\  (z_1,z_2,z_3) \to ({\ov z}_1,{\ov z}_2,{\ov z}_3)$. The model
contains four kinds of O6-planes, associated to the actions of
$\Omega R$, $\Omega R\theta$, $\Omega R \omega$, $\Omega R\theta\omega$.
We will focus on the open string (charged) spectrum. The closed string sector
contains gravitational supermultiplets as well as orbifold moduli
and is straightforward to determine. The cancellation of the  RR crosscap
tadpoles requires an introduction of $K$ stacks of $N_a$ D6-branes
($a=1,\ldots, K$) wrapped on three-cycles (taken to be the product of
1-cycles $(n_a^i,m_a^i)$ in the $i^{th}$ two-torus), and their images under
$\Omega R$, wrapped on cycles $(n_a^i,-m_a^i)$.

The rules to compute the spectrum are analogous to those in \cite{bkl}.
Consequently models with all tori orthogonal lead to even number of
families. Hence we consider models with one tilted $T^2$, where the
tilting parameter is discrete and has a unique non-trivial value \cite{ab}. 
This mildly modifies the closed string sector, but has an
important impact on the open string sector. Namely, a D-brane 1-cycle
$(n_a^i,m_a^i)$ along a tilted  torus is mapped to $(n^i,-m^i-n^i)$. It is
convenient to define ${\widetilde m}^i=m^i-\frac 12 n^i$, and label the
cycles as $(n^i,{\widetilde m}^i)$. 

The orbifold actions on the Chan-Paton indices of the branes, for each
stack of D6$_a$-branes, and their $\Omega R$ images, denoted by
D6$_{a'}$-branes, are as follows:
\beqa
&& \gamma_{\theta,a} =  \diag(i \id_{N_a/2},-i \id_{N_a/2}\, ; 
-i \id_{N_a/2},i \id_{N_a/2}) \nonumber \\
\nonumber \\
&& \gamma_{\omega,a}  =  \diag \left[ 
\pmatrix{0 & \id_{N_a/2} \cr -\id_{N_a/2} & 0 } \; ; \;
\pmatrix{0 & \id_{N_a/2} \cr -\id_{N_a/2} & 0 } \right] \nonumber \\
\nonumber \\
&& \gamma_{\Omega R,a}  = 
\pmatrix{ & & \id_{N_a/2} & 0 \cr 
& & 0 & \id_{N_a/2} \cr 
\id_{N_a/2} & 0 & & \cr 
0 & \id_{N_a/2} & & \cr }
\eeqa

The model is constrained by RR tadpole cancellation conditions. In
$\Omega R$ orientifolds twisted tadpoles vanish automatically
\cite{blumen,bonn}, whereas untwisted RR tadpoles require 
cancellation of D6-brane and O6-plane 7-form charges. For models with
a tilted third two-torus, they read
\beqa
\sum_a N_a n_a^1 n_a^2 n_a^3 - 16 & = & 0 \nonumber \\
\sum_a N_a n_a^1 m_a^2 {\tilde m}_a^3 + 8 & = & 0 \nonumber \\
\sum_a N_a m_a^1 n_a^2 {\tilde m}_a^3 + 8 & = & 0 \nonumber \\
\sum_a N_a m_a^1 m_a^2 n_a^3 + 16 & = & 0 
\label{tadpole2}
\eeqa
The solutions of the above equations define a consistent model; the
construction of the resulting spectrum is discussed in detail in
\cite{CSU}. Here we only summarize the results for D6-branes not parallel
to O6-planes (for zero angles, the spectrum follows from \cite{berkooz}).
The $aa$ 
sector (strings stretched within a single stack of D6$_a$-branes) is
invariant under $\theta$, $\omega$, and is exchanged with $a'a'$ by the
action of $\Omega R$. For the gauge group, the $\theta$ projection breaks
$U(N_a)$ to $U(N_a/2)\times U(N_a/2)$, and $\omega$ identifies both
factors, leaving $U(N_a/2)$. Concerning the matter multiplets, we
obtain three adjoint $N=1$ chiral multiplets. 

The $ab+ba$ sector, strings streched between D6$_a$- and D6$_b$-branes,
is invariant, as a whole, under the orbifold projections, and is mapped to
the $b'a'+a'b'$ sector by $\Omega R$. The matter content before any
projection would be given by $I_{ab}$ chiral fermions in the bifundamental
$(N_a,{\ov N}_b)$ of $U(N_a) \times U(N_b)$, where 
\begin{eqnarray}
I_{ab} = 
\left( n_a^1 m_b^1 - n_b^1 m_a^1 \right)  
\left( n_a^2 m_b^2 - n_b^2 m_a^2 \right) 
\left( n_a^3 {\widetilde m}_b^3 - n_b^3 {\widetilde m}_a^3 \right) \nonumber
\end{eqnarray}
is the intersection number of the wrapped cycles, and the sign of $I_{ab}$
denotes the chirality of the corresponding fermion ($I<0$ giving
left-handed fermions in our convention). For supersymmetric
intersections, additional massless scalars complete the corresponding
chiral supermultiplet. In principle, one needs to take into account the
orbifold action on the intersection point. However the final result
turns out to be insensitive to this subtletly and is still given by 
$I_{ab}$ chiral multiplets in the $(N_a/2,\overline{N_b/2})$ of $U(N_a/2)
\times U(N_b/2)$. A similar effect takes place in $ab'+b'a$ sector, for
$a\neq b$, where the final matter content is given by $I_{ab'}$ chiral
multiplets in the bifundamental $(N_a/2,N_b/2)$. 

For the $aa'+a'a$ sector the orbifold action on the intersection points
turns out to be crucial. For intersection points invariant under the
orbifold, the orientifold projection leads to a two-index antisymmetric
representation of $U(N_a/2)$, except for states with $\theta$ and $\omega$
eigenvalue $+1$, where it yields a two-index symmetric representation.
For points not fixed under some orbifold element, say two points fixed
under $\omega$ and exchanged by $\theta$, one simply keeps one point, and
does not impose the $\omega$ projection. Equivalently, one considers all
possible eigenvalues for $\omega$, and applies the above rule to read off
whether the symmetric or the antisymmetric survives. A closed formula for
the chiral piece in this sector is given in \cite{CSU}.

The condition that the system of branes preserve the $N=1$ supersymmetry
requires \cite{bdl} that each stack of D6-branes  is related to
the O6-planes by a rotation in $SU(3)$: denoting by $\theta_i$ the angles
the D6-brane forms with the horizontal direction in the $i^{th}$
two-torus, supersymmetry preserving configurations must
satisfy
$
\theta_1\, +\, \theta_2\, +\, \theta_3\, =\, 0
$.
In order to simplify the supersymmetry conditions within our
search for realistic models, we will consider a particular ansatz:
$(\theta_1,\theta_2,0)$, $(\theta_1,0,\theta_3)$ or $(0,\theta_2,\theta_3)$. 

Due to the smaller number of O6-planes in tilted configurations, the RR
tadpoles conditions are very stringent for more than one tilted torus.
Focusing on tilting just the third torus, the search for theories with
$U(3)$ and $U(2)$ gauge factors carried by branes at angles and three
left-handed quarks, turns out to be very constraining, at least within our
ansatz. We have found essentially a unique solution. The D6-brane
configuration with wrapping numbers $(n_a^i,\widetilde{m}_a^i)$ is
given in Table \ref{cycles3family}.

\begin{table} 
[htb] \footnotesize
\renewcommand{\arraystretch}{1.25}
\begin{center}
\begin{tabular}{|c||c|l|} 
Type & $N_a$ & $(n_a^1,m_a^1) \times
(n_a^2,m_a^2) \times (n_a^3,\widetilde{m}_a^3)$ \\
\hline
$A_1$ & 8 & $(0,1)\times(0,-1)\times (2,{\widetilde 0})$ \\
$A_2$ & 2 & $(1,0) \times(1,0) \times (2,{\widetilde 0})$ \\
\hline
$B_1$ & 4 & $(1,0) \times (1,-1) \times (1,{\widetilde {3/2}})$ \\
$B_2$ & 2 & $(1,0) \times (0,1) \times (0,{\widetilde {-1}})$ \\
\hline
$C_1$ & 6+2 & $(1,-1) \times (1,0) \times (1,{\widetilde{1/2}})$ \\
$C_2$ & 4 & $(0,1) \times (1,0) \times (0,{\widetilde{-1}})$ \\
\end{tabular}
\end{center}
\caption{\small D6-brane configuration for the three-family model.}
\label{cycles3family}
\end{table}
  
The $8$ D6-branes labeled $C_1$ are spit in two parallel but not
overlapping stacks of $6$ and $2$ branes, hence lead to a gauge group
$U(3)\times U(1)$. 
Interestingly, a linear combination of the two $U(1)$'s is actually 
a generator within the $SU(4)$ arising for coincident branes. This ensures 
that this $U(1)$ is automatically non-anomalous and massless (free of
linear couplings to untwisted moduli) \cite{afiru,imr}, and turns out to
be crucial in the appearance of hypercharge in this model.

For convenience we consider the $8$ D6-branes labeled $A_1$ to be away
from the O6-planes in all three complex planes. This leads to
two D6-branes that can move independently (hence give rise to a group
$U(1)^2$), plus their $\theta$, $\omega$ and $\Omega R$ images. 
These $U(1)$'s are also automatically non-anomalous and massless.
In the effective theory, this corresponds to Higgsing of $USp(8)$ down to
$U(1)^2$.

The open string spectrum is tabulated in Table \ref{spectrum3}. The
generators $Q_3$, $Q_1$ and $Q_2$ refer to the $U(1)$ factor within the
corresponding $U(n)$, while $Q_8$, $Q_8'$ are the $U(1)$'s arising from
the $USp(8)$. 
The hypercharge is defined as:
\beqa
Q_Y & = & \frac 16 Q_3 - \frac 12 Q_1 + \frac 12 (Q_8+Q_8') 
\label{hyper}
\eeqa
From the above comments, $Q_Y$ as defined guarantees that $U(1)_Y$ is
massless.
The theory contains three Standard Model families, plus one exotic
chiral (but anomaly-free) set of fields, and multiplets with vector-like
quantum numbers under the SM gauge group.

\begin{table} 
[htb] \footnotesize
\renewcommand{\arraystretch}{1.25}
\begin{center}
\begin{tabular}{|c|r|c|c|c|c|c|c|} 
Sector & 
Non-Abelian Reps. &
$Q_3$ & $Q_1$ & $Q_2$ & $Q_8$ & $Q_8'$ & $Q_Y$ 
\\
\hline
$A_1 B_1$  & 
$3 \times 2\times (1,{\ov 2},1,1,1)$ & 
0 & 0 & $-1$ & $\pm 1$ & 0 & $\pm \frac 12$ 
\\
          & 
$3\times 2\times (1,{\ov 2},1,1,1)$ &
0 & 0 & $-1$ & 0 & $\pm 1$ & $\pm \frac 12$ 
\\
$A_1 C_1$ & 
$2 \times (\ov{3},1,1,1,1)$ &
$-1$ & 0 & 0 & $\pm 1$ & 0 & $\frac 13, -\frac 23$ 
\\
         & 
$2 \times (\ov{3},1,1,1,1)$ &
$-1$ & 0 & 0 & 0 & $\pm 1$ & $\frac 13, -\frac 23$ 
\\
          & 
$2 \times (1,1,1,1,1)$ &
0 & $-1$ & 0 & $\pm 1$ & 0 & $1,0$ 
\\
          & 
$2 \times (1,1,1,1,1)$ &
0 & $-1$ & 0 & 0 & $\pm 1$ & $1,0$ 
\\
$B_1 C_1$ & 
$(3,{\ov 2},1,1,1)$ &
1 & 0 & $-1$ & 0 & 0 & $\frac 16$
\\
            & 
$(1,{\ov 2},1,1,1)$ &
0 & 1 & $-1$ & 0 & 0 & $-\frac 12$ 
\\
$B_1 C_2$ & 
$(1,2,1,1,4)$ &
0 & 0 & $1$ & 0 & 0 & 0 
\\
 $B_2 C_1$ & 
$(3,1,2,1,1)$ &
1 & 0 & 0 & 0 & 0 & $\frac 16$ 
\\
          & 
$(1,1,2,1,1)$ &
0 & 1 & 0 & 0 & 0 & $-\frac 12$ 
\\
$B_1 C_1^{\prime}$ & 
$2\times (3,2,1,1,1)$ &
1 & 0 & 1 & 0 & 0 & $\frac 16$ 
\\
                   & 
$2\times (1,2,1,1,1)$ &
0 & 1 & 1 & 0 & 0 & $-\frac 12$ 
\\
$B_1 B_1^{\prime}$ & 
$2\times (1,1,1,1,1)$ & 
0 & 0 & $-2$ & 0 & 0 & 0 
\\
                   & 
$2\times (1,3,1,1,1)$ & 
0 & 0 & $2$ & 0 & 0 & 0 
\end{tabular}
\end{center}
\caption{\small Chiral Spectrum of the open string sector in the
three-family model. The non-Abelian gauge group is
$SU(3) \times SU(2) \times USp(2) \times USp(2) \times USp(4)$.
Notice that we have not included the $aa$ sector
piece, even though it is generically present in the model. The non-chiral
pieces in the $ab$, $ab'$ and $aa'$ sectors are not present for branes at
generic locations, hence they are not listed here. 
\label{spectrum3}}
\end{table}


Even though the model is an explicit string realization of the
brane world scenario, the string scale is of the order of four-dimensional
Planck scale because Standard Model gauge interactions are embedded in
different D6-branes. The experimental bounds on Kaluza-Klein replica of
Standard Model gauge bosons imply that the internal dimensions cannot be
large \cite{bgkl}.

Quarks, leptons and Higgs fields live at different intersections, hence
the Yukawa couplings among the Higgs and two fermions arise from a string
worldsheet of area $A_{ijk}$ (measured in string units)
stretching between the three intersections
\cite{afiru}, $Y_{ijk} \sim \exp(- A_{ijk} )$. Note that one
family of quarks and leptons do not have renormalizable couplings with the
Higgs field, due to the uncancelled $Q_2$ charges,
and the only chiral multiplets which carry opposite $Q_2$ charges
are charged under the weak $SU(2)$. 

This model is supersymmetric for some specific choice of complex structure
moduli, which determine the angles satisfying the supersymmetry condition.
The supersymmetry breaking effect when the condition is violated is
reflected as a Fayet-Iliopoulos term for the $U(1)$ gauge fields. It is
proportional to the deviation from the supersymmetric limit, and
reproduces correctly the tachyonic scalar masses, and zero fermion masses.
The corresponding D-term is expected to be cancelled by vacuum expectation
values (VEVs) of the
tachyonic scalar fields, hence shifting the configuration to a corrected
vacuum, where some intersecting D6-branes are recombined, triggering
gauge symmetry breaking.

It is quite remarkable
that these chiral models are related to the non-chiral model of
\cite{berkooz}, by recombinations of the 3-cycles on which the D6-branes
wrap.
This is the T-dual of a $D=4$ version of the small instanton transition
\cite{small}.
Clearly, our search is in no sense exhaustive.
There exist different variants, in the framework descibed here, obtained by,
{\it e.g.},
(i) changing the additional branes not directly involved in the SM
structure, 
(ii) allowing branes rotated at an angle 
in all three tori, and (iii) compactifying Type IIA orientifold on
a different orbifold which preserves $N=1$ supersymmetry.
However, within our
ansatz for the angles in the case of $\IZ_2 \times \IZ_2$
orbifold,
the requirements of supersymmetry, Standard Model-like
gauge group and
number of chiral families are rather stringent, and the model presented 
here is relatively unique.
Let us note that D6-branes wrapping around supersymmetric 3-cycles
with three non-trivial angles, say $(n_i,m_i)=(1,1)\times(1,1)\times(1,-1)$,
contribute to some (but not all) tadpole conditions with the same
sign as that of an O6-plane -- a feature which is absent when
all the D6-branes are parallel to some O6-planes.
It would be interesting to explore such
variants to eliminate the additional vector-like matter and the
extra exotics. We leave this for further investigation.

\acknowledgments

We thank Gerardo Aldazabal, 
Savas Dimopoulos, Jens Erler,
Jaume Gomis, Luis Ib\'a\~nez, 
Paul Langacker, Raul Rabad\'an,
Matt Strassler for 
discussions.
M.C. and G.S. thank the Theory Division at CERN for
hospitality during the course of the work. A.M.U. thanks M.~Gonz\'alez
for kind encouragement and support.
This work was supported in part by U.S.\ Department of Energy Grant
No.~DOE-EY-76-02-3071 (M.C.), in part by the Class of 1965 Endowed Term Chair
(M.C.), UPenn SAS Dean's funds (G.S.) and the NATO Linkage grant 97061 (M.C.).

\end{document}